# LLMs Integration in Software Engineering Team Projects: Roles, Impact, and a Pedagogical Design Space for AI Tools in Computing Education

Ahmed Kharrufa[1, *], Sami Alghamdi[1], Abeer Aziz[1], Christopher Bull

[1] Open Lab, School of Computing, Newcastle University

[*] Corresponding author: ahmed.kharrufa@newcastle.ac.uk

**ABSTRACT**

This work takes a pedagogical lens to explore the implications of generative AI (GenAI) models and tools, such as ChatGPT and GitHub Copilot, in a semester-long 2nd-year undergraduate Software Engineering Team Project. Qualitative findings from survey (39 students) and interviews (eight students) provide insights into the students' views on the impact of GenAI use on their coding experience, learning, and self-efficacy. Our results address a particular gap in understanding the role and implications of GenAI on teamwork, team-efficacy, and team dynamics. The analysis of the learning aspects is distinguished by the application of learning and pedagogy informed lenses to discuss the data. We propose a preliminary design space for GenAI-based programming learning tools highlighting the importance of considering the roles that GenAI can play during the learning process, the varying support-ability patterns that can be applied to each role, and the importance of supporting transparency in GenAI for team members and students in addition to educators.

**Keywords:** Generative AI, GenAI, Large Language Models, LLMs, CoPilot, Team Projects, Computing Education, Pedagogy, Design Space, Software Engineering Education,

## 1 INTRODUCTION

The implications of generative AI (GenAI) in education are currently a very active area of research. Some researchers are exploring this in the context of education in general (e.g. [11, 49]), while others are focusing on specific disciplines. What is special about GenAI and computing education is the development of models (such as Codex [57]) and tools such as GitHub Copilot that specifically support code generation. As such, this topic has gained an even greater research interest in computing education recently (e.g. [17, 18, 20, 29, 37, 44, 53]). Such research ranged from evaluating the quality of GenAI generated code and its impact on assessment (e.g. [19, 38, 41, 51]), to many that looked at educators or experienced developers' perspectives (e.g. [6, 11, 15, 31]). Accordingly, it is critical to understand how these tools can align with educational objectives set by educational organizations, aside from their capability to meet varied personal needs of students and educators. There is a growing consensus that prohibiting students from utilizing these tools is not practical nor is it beneficial for students learning. Instead, the focus may shift towards understanding, adopting, and adapting these tools to better serve fundamental educational objectives in the long run [11, 17, 27, 41].

In this work, we aim to add to the community's understanding of the potential role and implications of using GenAI tool in software engineering. To narrow the scope further, we are particularly interested in looking at the students' perspective based on practical use within *team-based* formal education and analyzing it through a *pedagogical* and *technology design* lens. As highlighted by French et al. [21], many of the papers reporting on students' perception of GenAI in computing education are not based on actual, or long-term, use and experiences where GenAI is integrated into their modules. This can be a natural consequence of the recentness of such technologies, the duration it takes to make and approve changes to taught modules at universities, and the eagerness to contribute to the discussion on the topic before students are given the opportunity to use them extensively in formal education.

Moreover, the students' perspective of the role of GenAI on teamwork and its effect on team dynamics in computing (i.e., GenAI's use in computing teams beyond one student collaborating with AI) remains insufficiently studied, even though teamwork is a common practice in computing projects and in professional careers in computing. We also noticed that apart from a few exceptions, the results and findings of existing research are rarely explicitly discussed from a learning theories and pedagogy perspectives (e.g. [32, 37, 53]) or from a technology design perspective (e.g. [23, 26, 33]). Instead, more focus can be found in literature on implementing strategies and policies on the use of GenAI at an organizational level (e.g. [12, 36, 40]) or on informing teaching practices (e.g. [8, 13, 26]). While these are extremely valuable and are urgently needed, we argue that a similar attention needs to be paid to how such tools (or wrappers around such tools) can be designed to magnify their potential benefits and mitigate their potential negative implications (e.g. [48]).

In this paper, we report on findings from a survey and interviews with 2nd year computing science students after completing a software engineering team project module. In this module, students were explicitly given permission to use GenAI tools to help with the coding part of their 7-week software engineering team project. Through examining our data and our review of current work in this space, we have identified several areas where this work can make a useful contribution. These areas include reporting on real educational experiences in using GenAI in a software engineering project, emphasizing GenAI's impact on teamwork within such projects, looking at data through a learning theory and pedagogical lens, and finally looking at the design space that can be informed by this research.

Accordingly, this paper contributes to our understanding of LLMs integration into computing education from four different perspectives: **1)** *Longitudinal perspective*: Enriching our understanding of computing students' perspective on GenAI use within a software engineering module over a full semester. **2)** *Teamwork perspective*: Understanding the role and implication of GenAI on teamwork within a software engineering context. **3)** *Pedagogical perspective*: The application of learning theory and pedagogy lenses to examine and report on the collected data. And **4)** *design perspective*: A proposal of a preliminary design space for designing GenAI based programming learning tools/environments.

## 2 RELATED WORK

To situate our work, this section provides an overview of the rapidly evolving field of Generative Artificial Intelligence in education, particularly within computing.

### 2.1 GenAI in Education and Computing Education

GenAI is significantly impacting education, driving a transformative shift in teaching and learning practices. OpenAI's ChatGPT, as highlighted by Velibor Božić [8] and Michael Neumann [36], represents a significant advancement in GenAI technology with broad applications in education. Božić discusses ChatGPT's utility in language learning, writing assistance, automated grading, and personalized education. Similarly, Neumann's work emphasizes ChatGPT's transformative potential in education but also calling for its responsible use.

As discussed in the introduction, GenAI's specialized support for code makes the role of GenAI in computing science education of particular importance. Accordingly, the integration of GenAI in computing science education opens new opportunities for academic innovation, enhancing learning through programming assistance, content generation, and personalized instruction (e.g. [11, 15, 43]). The academic sector is keenly exploring the potential and challenges of GenAI, focusing on ethical issues, equitable access, maintaining academic integrity, and responsible use [12, 25, 36]. As GenAI technologies like code generators become more common, there's an emerging challenge in ensuring that students maintain a strong foundation in traditional programming skills. Research into GenAI's efficacy in aiding beginner programmers has revealed both their advantages and limitations. For instance, GPT 4 shows improvements over 3.5 which again shows improved accuracy over earlier models like Codex, yet issues in output formatting and language accuracy remain [41]. Future research aims at developing conversational techniques to better support students in troubleshooting and refining Large Language Models (LLMs) for more precise guidance, underscoring the need for balancing the benefits with responsible GenAI use in education. Moreover, integrating GenAI tools in computing education raises concerns about their impact on learning goals and teaching methods. As an example, a study involving 12 students and six instructors indicated GenAI tools' usefulness in programming and learning but highlighted risks of overreliance and plagiarism, stressing the importance of including practical usage training and AI literacy in the curriculum [52]. Further, Paul Denny and colleagues point out the limitations of using natural language for complex programming tasks, underscoring that proficiency in high-level programming languages is still a crucial skill for computing students [16].

## 2.2 Students' Experience Using GenAI

Several recent studies have investigated the integration and impact of GenAI in education, focusing on student experiences across a range of academic disciplines. However, several of these studies rely on speculations, opinions, and survey-based data. For instance, a study from Hong Kong [12] involving 457 students and 180 teachers used surveys to collect both quantitative and qualitative data. The aim was to identify the necessary requirements and guidelines for developing AI policies applicable in university teaching and learning contexts. A similar study [40] highlighted concerns over GenAI misuse and proposed a GenAI Ecological Education Policy Framework for a future integration with GenAI, placing a strong emphasis on data privacy and security measures. Further, a study by Chan and Lee [13] utilized an online survey to assess how adapting GenAI to meet diverse learning preferences across generations could integrate traditional and AI-enhanced methods to promote critical thinking. This research reached out to potential participants via bulk emails, garnering responses from 399 students and 184 teachers, and offering valuable insights into attitudes towards GenAI and its potential effects on education. These mentioned works predominantly use survey-based methods to explore AI's role in educational contexts. The reported students' experiences from our research are based incorporating GenAI into a software engineering module for an entire semester. This method provides a deeper understanding of GenAI from the students' perspective through firsthand experience and application, rather than through speculative or supposed impacts.

Additionally, research conducted among students has included design experiments and the introduction of novel tool designs. Paul Denny and colleagues [16] evaluated AI-generated educational content against student contributions, highlighting AI's potential as an auxiliary educational tool. Hassan Khosravi and colleagues implemented RiPPLE [28], where students were tasked with creating code examples during a weekly lab session. Through blind evaluations focused on correctness and helpfulness, they discovered that AI-generated resources were comparable in quality to those produced by students, though differences in syntax and length were noted. Another investigation by Kazemitabaar et al. [26] assessed the effects of AI coding assistants like OpenAI Codex on the education of introductory programming students. This study involved 69 novice learners, aged 10-17, split into two groups—one using the AI code generator and the other did not. The group assisted by AI completed tasks more efficiently and with fewer errors, underscoring the utility of AI code generators in reducing stress and boosting engagement for beginners.

The "Grounded Copilot" study by Shraddha Barke and team [5] aimed to understand programmers' interactions with GitHub Copilot. Recruiting 20 participants, including 15 from academia and 5 from the industry, they conducted a programming task with Copilot for an hour, followed by interviews. The session recordings were used to inform a grounded theory analysis. The findings suggest directions for optimizing future AI programming assistants for better user engagement and efficiency.

A similar studies [42] examined novice programming students' interactions with GitHub Copilot, involving 19 university students aged 18-11 enrolled in an introductory programming course. These students, observed in a controlled environment for 30 minutes, mirrored in-class coding conditions. This study shed light on the challenges faced by first-time users, particularly how unsolicited Copilot suggestions could be more obstructive than helpful. Thus, novices showed a preference for an interaction model offering help on demand. These insights highlight the need for AI coding tools tailored to novice programmers' specific needs. The results also offer insights for integrating these technologies into introductory programming education and enhancing their usability to better assist novices in overcoming programming obstacle [42].

Furthermore, at London Metropolitan University [15], BSc Games Programming students engaged with GenAI through two models, focusing on current AI developments and games. The students utilized GenAI tools like ChatGPT in their assignments, reflecting on GenAI's role in enhancing creative and critical problem-solving skill. This inclusion of GenAI not only enhanced their ability to generate creative responses and infer context but also facilitated the development of critical problem-solving skills. Feedback mechanisms, supported by literature on the timing and specificity of formative feedback, played a crucial role in shaping their learning experience. The outcomes of this educational approach were reflected in the marks awarded, which ranged from 54 to 73 percent, demonstrating the students' ability to creatively tackle assignments and critically reflect on the use of GenAI tools in development processes, all while navigating the challenges set within the freedom of their academic environment [15].

Our paper, in contrast, extends beyond the specific use cases and tools investigated in these studies, aiming to provide a comprehensive understanding of GenAI's implications in computing science education from an empirical perspective. We focus on enriching the understanding of student perspectives on GenAI's potential and implications based on actual use in a SE module. While the cited research contributes valuable insights into GenAI's role in educational content creation, programming assistance, and skill development, our use of HCI's user-centered, collaborative work, and design lenses offer a more holistic view, considering the impact of GenAI on educational practices, teamwork dynamics, and the design of learning tools in SE education contexts.

**2.3 Educators' Stance on Using GenAI in Computing**

As various studies have highlighted, the integration of GenAI into computer science education is reshaping the educator's role. Al-Hossami's research looks into applying Socratic questioning alongside AI, specifically evaluating GPT-based models like GPT-4 and GPT-3.5 for debugging tasks [3]. This method promotes independent problem-solving among novice programmers through guided inquiry, showcasing both the strengths and limitations of these models compared to human expertise. Simultaneously, Sam and Guo's study [31], conducted through interviews with 20 programming instructors from nine countries, examines the broader impact of AI coding tools such as ChatGPT and GitHub Copilot on computing education. It explores instructors' perspectives on incorporating AI tools into curricula, raising issues related to bias, ethics, and the authenticity of AI-generated work [31].

Furthermore, research by Smolensky et al., [47] involving 389 students and 36 educators in Australia and the US, reveals concerns about academic integrity and the need for assessment modifications to encourage critical thinking. While educators lean towards altering assessments to accept AI's growing presence, student reactions are mixed, with some worrying about the potential loss of creativity [47]. Similarly, another research by Petrovska et al [39] explored the feasibility of integrating GenAI thoughtfully into software engineering education through formative and summative assessments, addressing coding capabilities and argument construction.

These studies highlight the complex challenges of adopting GenAI in educational contexts, advocating for a balanced approach that combines traditional teaching methods with AI's innovative capabilities. This strategy, in turn, is expected to prepare students and educators for the continually evolving technological landscape.

**2.4 AI in Teamwork**

Exploration into the role of GenAI in enhancing teamwork is in its preliminary phases, with a nascent but expanding body of work just beginning to uncover its potential benefits and implications. Zhang et al. investigated communication strategies within human-AI teams, pinpointing the necessity of proactive and balanced communication for effective collaboration [54]. The deployment of ChatGPT-4 in engineering education emerges as a promising strategy to improve team feedback mechanisms, striving to foster a nurturing and growth-oriented team atmosphere [46]. Moreover, GitHub Copilot's integration into software development routines signifies a transformation, redefining coding as a collaborative, rather than an isolated activity [6]. Comparative studies on human-AI versus traditional human-human pair programming shed light on AI's ability to bolster human efforts in educational frameworks, emphasizing the critical need for flexible and ethical adoption strategies [34]. These insights, while enlightening, underscore the urgent need for more extensive research to thoroughly understand and harness AI's, particularly GenAI's, capacity for facilitating teamwork in various fields and contexts.

In the context of qualitative analyses collaboration, Gao et al. developed CoAIcoder [22], a tool specifically designed to enhance human collaboration in qualitative analyses of data. This was achieved by leveraging AI to provide real-time code suggestions. The study concludes that AI can be effectively integrated into the inductive coding process, thereby promoting collaborative interpretation.

In engineering and computing education, teamwork emerges as a pivotal skill, with students and educators alike navigating the complexities of collaborative learning environments. A study was conducted to improve teamwork in first-year engineering design teams by utilizing GenAI, specifically ChatGPT-4, for feedback synthesis. Their research acknowledges the instrumental role of teamwork, underscored by the traditional use of tools like CATME for team formation and peer feedback [31].

However, most research in AI teamwork in computing education has focused on the interaction between humans and AI in pair programming settings. There is a need for more research looking at the role that GenAI can play as a member of a programming team. For instance, Barke [5] explored the Human-AI pair in the context of GitHub's Copilot and highlighted the shift in coding experience, emphasizing the growing importance of reviewing written code and discussing potential improvements in AI pair programming. Similarly, Ma et al. [34] investigated the dynamics of human-AI pair programming through literature, contrasting it with traditional human-human pair programming. The research suggests that while AI partners like GitHub Copilot can complement human programmers by managing tasks like code generation, there are unique challenges and opportunities in optimizing this collaboration for educational contexts and professional development. These studies, while insightful, underscore the broader need for extensive research to fully understand and leverage AI's and more specifically GenAI potential in teamwork across diverse domains.

Our paper, on the other hand, sets itself apart from those mentioned in that it looks more closely at GenAI impact on teamwork within software engineering teams.

### 2.5 Pedagogical and Technology Design Perspectives in GenAI Research

Our review shows that only a few studies explicitly incorporated pedagogical or learning theory perspectives in examining the role of GenAI in computing science education. Interestingly, most studies that adopted these perspectives did so by developing their own tools to enhance existing GenAI technologies for educational purposes (e.g., [23, 26, 33]).

A significant line of inquiry has focused on assessing the impact of GenAI tools on learning, especially regarding the potential for overreliance on these technologies. Kazemitabaar et al. [26] created a unique Integrated Development Environment (IDE) that leveraged the Codex LLM to aid novice programmers aged 10-17 in their learning processes. Their findings indicated that GenAI tools enabled newcomers to programming to code more efficiently and effectively, reducing frustration. Crucially, reliance on GenAI did not diminish the students' ability to modify code manually when GenAI support was absent. The study concluded with three key design implications for GenAI tools: **1)** enhanced support for absolute beginners, stemming directly from the study's target demographic; **2)** mechanisms to control overuse, preventing dependency that could hinder learning progress; and **3)** improved support for constructing effective writing prompts.

Similarly, Liffiton et al. [33] developed CodeHelp, a GenAI-powered tool designed with 'guardrails' to offer on-demand assistance to programming students without directly providing solutions. These guardrails aim to curb students' over-reliance on GenAI, offering an innovative way to integrate GenAI tools into learning while enabling educators to monitor and analyze student engagement with the technology. The evaluation of CodeHelp was positive, with students appreciating its support in completing tasks successfully and enhancing their learning. Key themes from student feedback included error correction, support for independent learning, speed, and improved understanding. A noted concern was the tool sometimes presenting content not yet covered in the course. Although specific design implications were not detailed, the study suggests valuable design considerations like implementing guardrails, providing educators with dashboards for monitoring GenAI usage, and ensuring GenAI-generated content aligns with course material.

Another research focus is the pedagogical value of learning from worked examples, a widely studied approach in education, particularly in programming education [1, 11, 55]. Leinonen et al. [32] examined how code explanations by students compare with those generated by GenAI in terms of accuracy, length, and understandability. They found that while GenAI and student explanations were similar in length, GenAI-generated explanations were perceived as more accurate and easier to comprehend. This suggests GenAI's potential in producing resources for students, especially beneficial in early learning stages where example-based learning is common. Acknowledging the educational advantages of worked examples and the time required to create quality ones, Jury et al. [23] introduced 'WorkedGen' for generating interactive worked examples. Their study involving around 400 first-year Python programming students demonstrated that GenAI could create effective, high-quality examples that students found valuable, particularly due to the interactive possibilities offered by GenAI.

## 3 METHOD

### 3.1 Study Context

This study focused on students enrolled in a 2nd-year undergraduate Software Engineering Team Project module at a UK University, a compulsory course for all Computing majors. This module spans seven weeks and is the only module the students work on during this time. It emphasizes software engineering principles, involves extensive programming, and it is designed as a key team-based project coursework. Programming activities commence in the second week or earlier, allowing students to devise their projects around the United Nations Sustainable Development Goals[1], typically leading to the creation of mobile or web applications based on a multi-tier architecture. While the module centers on software engineering, it also adopts a comprehensive approach by integrating Human-Computer Interaction (HCI) elements.

---

[1] https://sdgs.un.org/goals

At the outset of the second week—following an initial week dedicated to team formation and the definition of project ideas and requirements—the module leader informed the students about the availability of GenAI tools for their projects. This introduction covered a range of GenAI technologies, including ChatGPT, Google's Bard, and GitHub Copilot, emphasizing the students' freedom to explore and optionally utilize these tools in their project development. However, a significant condition was set: these tools were permitted exclusively for project creation and code development, not for composing the assessed textual design report. This distinction was crucial as the evaluation criteria focused on the application of software engineering principles over algorithmic quality. The students received repeated reminders about the use of these tools and the associated constraints throughout the module, ensuring clarity on the guidelines and mitigating potential issues such as plagiarism or other assessment discrepancies. This was particularly aimed at maintaining awareness among all students, including those potentially less engaged, who might overlook verbal and written announcements.

Ethical approval was sought from the host academic institution and awarded under reference #31848/2023.

### 3.2 Data Collection and Analysis

A mixed-method approach was adopted for this study to capture both comprehensive and detailed insights, utilizing a combination of a survey and follow-up in-depth semi-structured interviews. The survey was conducted initially to gather a general understanding of the students' perceptions regarding their use of GenAI tools in their projects. At the survey's conclusion, participants had the opportunity to provide their contact details if they were interested in further discussing their experiences through a semi-structured interview. This second phase was designed to go deeper into the survey findings, offering a platform for more detailed and nuanced insights.

Thirty-nine students consented to participate in the survey phase of our study. Within this cohort of survey participants, a demographic breakdown reveals a gender distribution wherein 77% identified as male, 18% as female, 2.5% as transgender, and 2.5% as non-binary. The cultural background of participants exhibited considerable diversity, with White British identified as the predominant category.

Following the survey, eight students agreed to partake in the subsequent in-depth interviews, with seven identifying as male and one as female. Although a consistent interview guide was utilized across all sessions, the semi-structured nature of each discussion led to unique insights, because of the dynamic interaction between the interviewer and each participant. This method was particularly effective given the varied national and cultural backgrounds of the student body, enabling the collection of rich and diverse data that accurately reflected the personal contexts and experiences of the participants.

#### 3.2.1 Overview of survey and interview questions

To gain a well-rounded understanding of students' experiences with GenAI tools in their team projects, we developed surveys and interviews to complement each other. The survey questions (Table 1) aimed to gather wide insights into the integration of GenAI tools into coursework, their influence on programming skills, team dynamics, and the overall learning experience. Our survey combined multiple-choice options, Likert scales, and open-ended questions to gather qualitative data.

Table 1: Summary of survey questions

| Question Group | Description |
| --- | --- |
| Personal Information, Environment and Conditions of AI Usage | This section included seven questions asking about: group number within the module, gender, cultural background (optional), role in the team, the AI tool used for the software engineering project, self-rated programming skills before using the AI tool, and the frequency of AI tool usage within the team. |
| Overall Experience and Impact of AI tools | This section contained six questions evaluating the experience of using AI for coding, rated from very negative to very positive; the extent to which AI tools were helpful in writing code; reasons for the ratings provided in the previous question; the influence of AI tools on approaches to programming; confidence in coding abilities after using AI tools; and agreement with the statement that AI tools promote critical thinking and problem-solving. |
| Details of experiences with AI Tools | This section contained eight questions covering: agreement with AI tools facilitating efficient completion of programming tasks, impact of AI tools on the project, assistance from AI tools in learning coding, likelihood of recommending AI tool usage in other programming courses, preferences for AI tool utilization in programming modules, comfort level with using AI systems in assignment assessment, and elaboration on comfort levels with AI assessment. |

The interview questions (Table 2) aimed to provide an in-depth exploration of these subjects, additionally examining the impact of GenAI integration on team workflows and learning, and the anticipated future role of AI in the software engineering discipline.

Table 2: Summary of interview questions

| Question Group | Description |
| --- | --- |
| Overall Experience | A warm-up question: "Could you share your thoughts on using AI tools?". |
| Integration and Impact of AI in Workflow and Learning | This category includes questions on how AI is integrated into the software development workflow, whether tools like ChatGPT and GitHub Copilot assist in understanding new programming concepts or approaches and how, and how the use of AI has altered the pre-AI help-seeking process, including inquiries to peers, using Stack Overflow, and general online search. |
| AI and Team Dynamics | Includes questions on the team's use of AI and its impact on the team's progress, as well as whether any changes were observed in the team's decision-making process for coding tasks due to the use of these AI tools. |
| Trust and Verification in AI-Generated Code | Includes a question on whether participants trust the AI-generated code enough to use it as is, or if they verify it first. It also probes whether their trust in these tools has changed over time and, if so, how. Additionally, it inquires if participants have used AI tools to test their code and, if so, in what ways. |
| Assessment Using AI Tools | Includes questions on whether the potential use of AI tools for assessment by instructors would alter students' approaches to completing assignments. It also asks students whether they believe the use of these AI tools offers a fair and accurate representation of their coding skills, along with reasons for their perspective. |
| Ethical Implications of AI in Coding | Includes questions on whether participants had any ethical concerns while using GenAI tools for programming tasks. It also seeks participants' views on the potential risks and benefits concerning the ethical use of these tools. |
| Future of Software Engineering Education and Roles | Includes questions on participants' perceptions of the potential impact of GenAI tools on computer science education and its effect on preparing students for future software development roles. It also explores whether participants believe students taught with or allowed to use these tools will have advantages or disadvantages upon graduation compared to those who were not. |

*3.2.2 Analyses*

To uncover the nuances of students' interactions with their selected LLM GenAI tools, a reflexive thematic analysis approach [9, 10] was utilized to analyze the survey and interviews. This analysis aimed to understand how students experience LLM tools as a source of support in their SE team project. We followed an iterative, deductive-inductive process that began with two researchers examining all the data through the broad lens of teamwork, scaffolding and fading, critical thinking, and problem-solving. Codes and sub-themes were created and reviewed by these two researchers, then shared with the larger research team. After achieving a full understanding of the data and identifying the intended contribution of this research, the themes and subthemes were further refined by the research team in one additional iteration.

## 4 RESULTS

### 4.1 Quantitative Insights From Survey

We provide a summary of the survey results and some key insights, as the survey was used to inform the interview design. The full survey questions and results are provided in the supplementary materials.

The 39 respondents used a variety of GenAI tools (some using multiple tools), including: ChatGPT (33), GitHub Copilot (15), Google Bard (2), Microsoft Bing Chat (1), and 3 did not use any GenAI. When asked a related question about "How often were you responsible for using AI in your team?", we find a spectrum of usage (Figure 1) with most respondents occasionally using GenAI (44%).

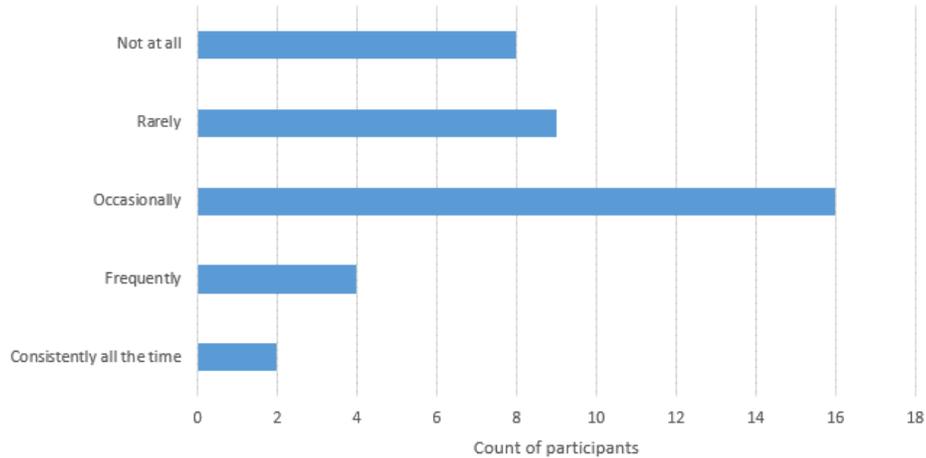

Figure 1: Frequency of GenAI use.

Generally, the respondents had a slightly positive experience with GenAI (Q8, 3.64 average) and, in their view, its helpfulness was neutral in assisting writing code for their projects (Q9, 3.08 average)—see Figure 2. However, the respondents were slightly more likely to recommend these GenAI tools to be considered for other courses.

The students were also asked to consider if the GenAI tools promoted critical thinking and problem solving (Q13), and whether the tools enabled them to complete the programming part of their work more efficiently (Q14) Figure 3. There is a clear spread of views, though in terms of critical thinking and problem-solving enhancement attributed to these AI tools, the majority, accounting for almost 53%, either agreed or strongly agreed. Although when asked if they agree or disagree that GenAI tools enabled them to "complete the programming part more efficiently", 77% of respondents either chose "strongly agree" or "agree"; 18% of the respondents perceived that these tools sometimes produce less than ideal code or do not yield comprehensive solutions, requiring subsequent edits.

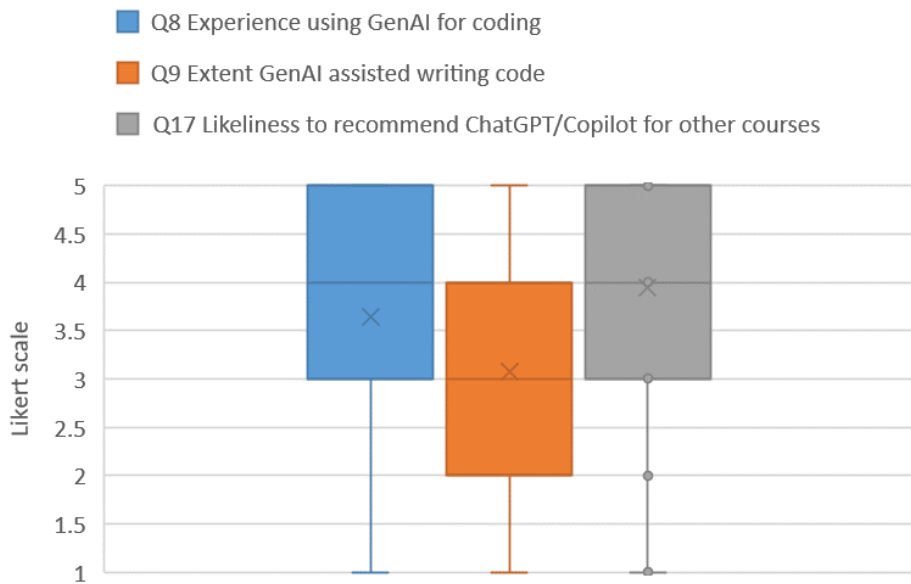

Figure 2: Likert scales of experiences with GenAI next to how helpful students perceived GenAI. All used a 5-point Likert scale. Q8 scale was labelled "Very negative" to "Very positive". Q9 scale was labelled "Not at all helpful" to "Extremely helpful". Q17 scale was labelled "very likely" to "very unlikely".

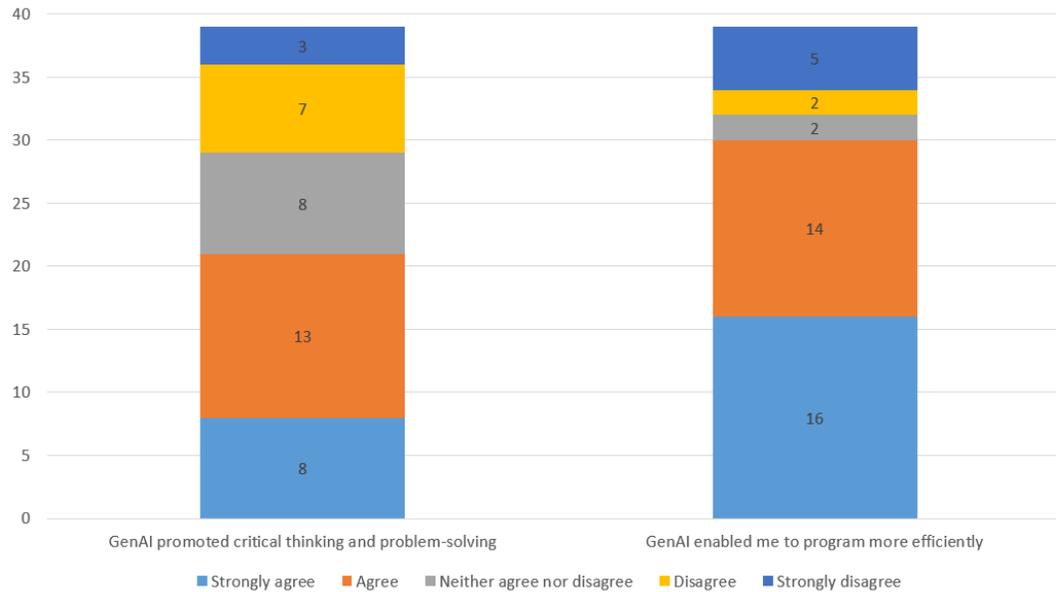

Figure 3: Students' indication whether they agree or disagree that GenAI tools promote critical thinking and enable efficient programming.

In the following qualitative results, we did not include speculations from students who selected 'not at all' or 'rarely' in the survey results when answering the question about their frequency of use of AI unless this data provides real experiences as opposed to speculations. For those who took part in interviews, we focused on their richer elaborated answers from the interviews rather than in the survey. All survey responses start with **S** and all interview responses start with **I**.

The results are grouped into the following themes: Teamwork, supporting learning, challenges to learning, and coding specific usage,

**4.2 Teamwork**

The data highlights a generally positive impact of GenAI on team confidence, dynamics, and productivity. It also shed light on some of the difficulties that can result from using GenAI in a team setting.

*4.2.1 Bridging the skills gap and increasing the team's confidence.*

Some students viewed AI tools as enablers that bridged the skill gap between team members, thus allowing all team members to contribute effectively to the project (e.g. "*I think it made everyone seem to have an increased level of confidence, I think, in what level we could all complete the project…everybody has vastly different skill sets and it can sometimes be hard to manage that…But now that the AI is a thing, I feel like that gap [in experience in React] can now be bridged, because if you understand how to utilize the AI in an effective manner, it doesn't hinder you as much and you can still have a large contribution to the project…*", I1).

Such bridging of skills gaps led to higher level of confidence and team-efficacy (e.g. "*We were a little bit disheartened when we found out that one of our members wasn't going to do anything. But we were like, Oh, look, it's not that bad, especially with the AI tool now, we've got an idea of how to do this. So we don't really need them.*", I2). The overall effect was seen to bring a sense of calmness to teams believing that they can solve more challenging problems than before ("*I definitely noticed a difference as the project manager. People remained calm and knew that with ChatGPT, they could resolve errors or come up with new functions. It brought a sense of calmness, and they had confidence that ChatGPT could help solve any problem.*", I3).

*4.2.2 AI as a team member*

One mechanism in which GenAI contributed to such increased confidence and calmness is by playing the role of an additional team member whose contribution to the team can range from (1) <u>doing tasks</u> that other team members normally do, (**2**) to <u>provide</u>

more capacity such as reviewing documentation and code written by others (e.g. "*It's also helpful in stages where you have to review things like a technical design document. Instead of always asking a team member to review what you've written, it might be quite useful to quickly paste it into ChatGPT and get some feedback to see if it's okay or not.*", I4) , and (**3**) to supporting the team in idea generation and brainstorming as repeatedly reported by the students (e.g. "*…when we met with my team for brainstorming, we used GPT to generate some project ideas. This was useful, as most people would propose simplistic ideas. But ChatGPT gave us quite a few unique ideas to discuss. Eventually, we decided on one of the ideas ChatGPT suggested…*"**,** I4; "*It was also used in conjunction with us, trying to think of ideas itself. So we sort of used what like AI had said to influence what we were thinking.*", I5; and "*helps generally with brainstorming ideas.*", S19).

In those examples, students treated GenAI as a collaborator that can influence their decision and contribute to their work as a peer rather than just a support tool doing certain mundane tasks.

*4.2.3  Improved the productivity of the team shifting its focus to the bigger picture, team skills, and testing*

The different roles and levels of support that GenAI contributed to the team (as detailed in previous and upcoming sections) increased the teams' productivity giving them more space and capacity to focus on more the bigger picture behind the project (e.g., "*Now that AI was open to use, it helped us turn our small ideas into code. I could focus on the bigger ideas and didn't have to worry about the time-consuming little syntax and stuff. I could just focus on the big picture and start programming. I feel like it allowed our group to create a high-quality project compared to if we weren't allowed to use AI.*", I3; and "*I think it was mainly used within the team to Build up... Boilerplate code... I think it just allowed us to focus more on the problems and developing features, things that you know, things that were important to us, and that we thought needed more development time*", I6).

Similarly, other students reported on how such increased productivity allowed them to focus on documentation and team skills (e.g. "*They [the AI tools] have given us more time to focus on the development of documentation and team skills, which were core components of this module.*", S27), and design and testing (e.g. "*It has reduced the time needed for coding, allowing us to focus responsibility on other important issues such as design and testing.*", S7).

Improving productivity, either directly or indirectly, emerged as a prominent subtheme, drawing numerous comments from both survey and interview responses.

*4.2.4  Difficulties caused by GenAI usage in the group*

Experiences of using GenAI at the team level were not always positive. One of the students who was among the most positive about the use of AI in computing, talked about the challenges caused by using AI at the team level including team members generating buggy code that they do not understand and treat as a black box which can negatively impact the team as a whole ("*as the one managing everyone's code and branches, it was sometimes difficult to understand the code. People might have generated code with AI without fully comprehending it and treated it like a black box, saying, 'It's working, here it is.' So, I had to debug a lot of code that some people might not have understood or even created themselves, if that makes sense.*", I3).

The same sentiment was expressed by others in the survey, that team members' use of AI resulted in subpar code that impacted the whole team especially when it was not clearly communicated that the code was GenAI generated and other team members had to spend extra time fixing such code (e.g. "*I found that often times the code provided by them [other team members] was not representing their skill set correctly and couldn't wrap my head around the mistakes in it, as I wasn't informed they're using AI. It slowed down our development and I don't think it helped anyone with their coding skills. I spent hours fixing code that was written by AI.*", S6; and "*Overall I believe they hindered my project, as many of my team members misunderstood how to utilize it properly. This resulted in subpar code that made no sense being present in the project that required amending later on*", S12).

### 4.3  Supporting Learning

"*I use AI a lot now every day, and I don't know how universities, schools, and the education system can continue as it is now. Almost every student, like 9 out of 10, is using AI and they can pass any exam and do anything with AI.*" (I3)

As demonstrated by this quote, the current and potential impact of GenAI on students learning and learning experience cannot be overstated. We wanted to use a pedagogical lens to look at students' data from the survey and interviews. Through our iterative analysis while familiarising ourselves with the data we found scaffolding to be a very common theme which can be further broken down into scaffolding using worked examples and scaffolding to provide a starting point. The other subthemes within the learning theme are supporting critical thinking and particularly learning about alternative techniques to solving the same problem and increased self-efficacy (at the individual level).

*4.3.1 Scaffolding*

In addition to playing a role as a team member, GenAI played an <u>educator</u> role, providing scaffolding to support students in their learning and problem solving. This was a very popular theme with many examples of how AI supported the students allowing them to do things they could not have done by themselves without support - normally from teaching staff. One student gave an excellent example of how AI helped them stretch themselves within their Zone of Proximal Development (ZPD) without even knowing about this concept ("*Really have improved it as without the use of AI i wouldn't have stretched myself as far it allowed me to go beyond my normal comfort zones as i knew i had some help if i got stuck."*, S22).

Other students gave perfect examples of scaffolding and fading again without knowing about the concept or being explicitly prompted about it (e.g., "*I personally find AI tools very useful for programming, especially for software engineers. They can write programs for you, providing a steppingstone and a boost. For instance, they can help you grasp the basics, but beyond that, as things get more advanced, their assistance diminishes.*", I7; and "*It's like an assistant that can provide necessary help, and if used properly, they [students] can learn how to do it themselves, in the most efficient way, and won't need the AI tools anymore.*", I2).

There are also many examples that talk about GenAI providing explanations and support in problem solving, frequently comparing it to human support. Below are just some of these examples:

"*…when I didn't fully understand some of the logic schemas, we needed to design to explain our application's logic. The lecture presentation was okay, but I needed a bit of extra information. ChatGPT was helpful here… If I don't fully understand something, then I might use GPT to explain it to me or to provide me with ideas.*", I4)

"*I now use it as a mentor. Normally, before AI, I would search online, use Stack Overflow, or ask a demonstrator or lecturer if I had specific problems. But now, I can provide ChatGPT with the context of the module and ask questions as if I'm speaking to an expert. I don't have to rely on real-life interactions. I can treat ChatGPT as an expert itself, and it has the context to provide me with equal or even better help than speaking to a TA or someone else.*", I3).

"*It made me more relaxed when approaching problems, as I had a helper (AI) available to me at all times.*" (S12). This comment is interesting as the same student also commented that "*Whilst generative AI gave me a basic idea of how to achieve a certain task, it was often inconsistent with its answers. The answers were only useful a fraction of the time, and it gives wrong answers with full confidence.*" As such, even with such a rather negative view of the outputs from GenAI, it was still seen as a useful helper that is always available.

*4.3.2 Scaffolding: Learning from worked examples.*

In our literature review, we looked at some work that explored the use of worked examples in programming education (e.g. [1, 55]) with [32] explicitly looking at GenAI's potential in this space. In examining our results, we identified repeated cases where the AI generated code is studied by the students as examples to learn from (e.g. "*It [Copilot] helps me improve my coding skills by providing better examples.*", I4; "*I think if you take the time to try and understand the code that it's generating, and then even asking it to explain line by line what it's doing, you can develop a really strong understanding of exactly what is going on in each function and each file.*", I1; "*While you could simply copy and paste the code AI provides, to truly understand what it does and how to use it in future projects requires you to research the methods and classes it provides. Therefore, I think it is quite useful for learning… By asking AI to generate code, you need to understand it before you can implement it, which means learning new methods. This can solidify the knowledge in your mind.*", I7; "*Makes learning easier by giving examples, and being able to describe what the code is doing*", S5; and "*Seeing the solutions and remembering what was suggested has taught me in the same way a teacher would.*", S16).

*4.3.3 Scaffolding: getting started*

Another type of scaffolding mentioned by a few students, which builds on the previous two, is "*help getting started*" (S4) on the project, or even a coding task. Depending on how this support is provided, this can fall under the 'partial-worked example' or 'completion examples' category, but in other cases it can be guidance or tips rather than code. Students repeatedly talked about GenAI providing help getting started and sometimes referring to its potential downsides (e.g. "*They give a place to start learning* *if the student is completely lost in where to begin, but other than that I don't believe they are much use in assisting learning as it is too easy to mindlessly use the code without understanding it.*", S23; "*Often if i wasn't sure how to start coding something i could ask chat gpt what it would suggest to code a certain a feature and it might suggest a library or two and some potential code however it was very rare it was good high quality working code.*", S22; and "*I think it helps some members of the group with a starting point, but it leads to far too much debugging to get it to work*", S17).

*4.3.4   Critical thinking and learning different ways of doing things.*

In addition to scaffolding, another important potential support for learning is developing students critical thinking skills. Only one student explicitly commented on the support of GenAI for problem solving and analytical skills ("*Programming generally builds up your problem-solving skills over time. If you're using AI properly, it's also going to build up your critical analysis skills. So, it's kind of a different skill that is being built up... If you're always thinking, 'What does this actually do?' when you look at code, and that's how AI is being taught in universities, then it could realistically enhance problem-solving skills*", I8). However, the most recurring comment in this space is that of helping students learn different ways of doing things which consequently widens their perspective and improve their critical thinking and analytical skills. The comments were very clear on the perceived positive impact of such alternative methods, and thus we are presenting a few of these quotes to reflect the students' emphasis on this point:

"*It can suggest different ways of doing things which you might not have thought of. Usually, it's not more efficient, but sometimes it is, and it can expand your knowledge through suggesting new syntaxes and new algorithms, etc., which might not have been in your comfort zone. It gives you the opportunity to expand and explore these hacks and codes and so on.*", (I8).

"*Using GitHub Copilot, I saw other methods - even quicker, more efficient methods to do what I wanted to do*" (I2).

"*I would use it to see if there are any alternative solutions to my problem.*" And "*They allowed me to find new approaches for completing tasks.*" (S9).

"*It sometimes also provides alternative point of view in implementing a function.*" (S18).

"*I learnt a lot about coding because often AI would suggest a more efficient way of writing code than the way I have wrote it by using new methods that I did not know.*" (S10).

"*These tools have used ways of coding in which I hadn't considered meaning that I have actually learnt new techniques and common practices through using AI tools.*" (S25).

*4.3.5   Self-efficacy*

A theme implicitly present in many comments was that the use of GenAI tools empowered some students to attempt more ambitious solutions and increased their confidence in their capabilities (e.g., see the teamwork section). More explicit references to this effect at an individual level were found in the survey, such as, "*Using AI made me less afraid of trying to implement more ambitious features to our project. I would also usually assume that the development time would significantly shorter as I would spend less time on writing boilerplate code.*" (S24) and "*I feel much more confident and can now program a much wider range of things*" (S29). These findings are particularly significant for programming education, as self-efficacy is often cited as a barrier to learning programming across all educational settings, both formal and informal [2, 24]. Interpreting this through the lens of Albert Bandura's self-efficacy theory[4], which posits that mastery experiences - be it successes or overcoming challenges - are the most potent sources of self-efficacy, these experiences reinforce individuals' belief in their abilities and motivate them to persist through difficulties. The examples provided show students reporting successful experiences with GenAI tools in their coding tasks, which boosted their confidence and willingness to tackle more challenging tasks. This underscores the importance of exposing students to experiences that promote feelings of empowerment and capability [45, 56], thereby positively influencing their self-efficacy. Such self-efficacy plays a crucial role in persistence and resilience against challenging tasks like coding, and GenAI tools appear to be effectively capable of supporting this feeling, indicating an important design direction that focuses on this affective element.

**4.4   Challenges to learning**

Despite the reported excitement about AI, students still realize its potential negative impact, or challenges on learning, if not used responsibly. In this theme, they talked about how the use of GenAI can hide the actual skills of the students as well as the potential negative impact on learning due to over reliance on the technology.

*4.4.1 Hiding students' real skills*

While talking about the challenges of GenAI when it comes to assessing students' work when they may have used GenAI, an interesting alternative way of looking at the same problem was raised by the student and that is the use of GenAI 'hides' the real skills of the students. This was a point raised by several students such as: "*I found that often times the code provided by them [team members] was not representing their skill set correctly*" (S6); "*sometimes, like, your ability could be hidden behind what the AI's coded. Like, you could use the AI to boost what you can do. It's more if you could understand what is doing, and AI can be furthering for the understanding.*" (I5); and "*If you're referring to general flow control, project implementation, then AI tools do reflect people's skills because they are planning and implementing projects based on their ideas. But if you're referring to the efficiency of code and the ability to write proper, optimized code, then I don't think AI tools fully reflect those skills.*" (I3).

One of the interviewees discussed how such hiding of skills does not only impact assessment and teamwork, but also the learning of the students' themselves. This results from the reduced chances of learning from mistakes which normally are very important learning opportunities for the students. When GenAI quickly corrects such mistakes without much deliberation on the side of the student, these opportunities go unnoticed and thus unrealized ("*I feel like some people may seem better than they actually are, making fewer mistakes. They might not learn from those mistakes because AI is fixing them.*", I2).

*4.4.2 It depends on the students' ability levels and usage patterns*

Most importantly, students' comments show that they mostly realize that eventually, it all depends on how they use AI and at what stage in their learning process. They emphasized that students need to learn the fundamentals first before starting to use AI to ensure it does not negatively impact learning (e.g., "*The use of AI is likely to increase, especially in the real world. I think it's useful for students, but they should still learn the basics without the use of AI. Without a solid understanding of the basics of a programming language, AI cannot provide much help.*", I7; and "*Being able to use AI did not affect me because my ability to code comes from understanding what im doing and I don't think AI is very helpful if you don't know what you're doing.*", S33).

Students also expressed the view that those who understand the fundamentals of programming and know how to use AI will be in an advantageous position compared to those who do not (e.g., "*Those who use AI to learn will definitely have an advantage compared to those who just use it to coast by. So, in all cases, there is an advantage. In the short term, everyone benefits because AI does the work. However, in the long term, those who use AI properly for learning and verify their code will have a massive advantage.*", I3; and "*In my opinion, those who use AI tools may have an advantage, provided they find the right balance between using these tools and their own cognitive abilities…However, if there's no limitation on how much they're used, students may face reality shocks as they often have to adapt to different situations and use their own thinking in the real world. So, if taught properly, those with a sound understanding of how to use AI tools will have an advantage. Otherwise, they might be at a significant disadvantage.*", I4).

While last comment refers to 'how' GenAI is taught at university, the following refers to 'when' it is taught ("*I think it depends how early on you're allowing them to use it. If it's too early on and they haven't picked up the skills yet, I think it actually could be a downside. On the plus side, if it was introduced a bit later, like in the software engineering project when people generally have a good understanding of programming, I think it'd be beneficial, really.*", I6). For others, 'it really depends on the individual' as GenAI can support learning at all stages if used responsibly ("*For novice learners, I feel like it's a good start. But if it's used too much, they could end up not learning how to code, because if they end up just relying on this, then it might not be very good. Because sometimes you don't have these AI tools… If someone is willing to learn, they will use AI responsibly and see it as a method to learn, almost like another textbook.*" (I2).

Along the same lines, and a slightly more concerned view was expressed by several of the survey students who mentioned that they have either not used GenAI at all, or used it 'rarely' as they worried that use of such tools would negatively impact their learning (e.g. "*I don't want to fully depend on a tool while still learning to code.*", S19; "*If I had used I feel that maybe I would've learned less*", S17; "*I did not use them but if I had, I think they'd have hindered my learning rather than assisted it. I'd just be copy and pasting rather than actually writing code, so I wouldn't understand it or be able to write it myself in future*", S32; and "*I feel using AI does not help with learning. It helps with fixing solutions faster but I don't believe you learn to code you just learn how to manipulate the ai to give you want you need*" (S33)

While the above comments were speculations, a demonstrable potentially negative impact is seen in the following quote "*with AI, we were coding at a higher level, using functions that we might not fully understand. We were editing the code on the fly and didn't have a specific way to verify it*" (I3).

### 4.5 Coding specific usage

While most of what has been mentioned earlier can be generalized to any discipline, some issues were more specific to coding such as writing code vs understanding its underlying concepts, the problem-solving process which typically relies a lot on sites such as Stack Overflow, code testing and bug fixing.

#### 4.5.1 Impact on what is learned and how.

At a high level, some students talked about GenAI's impact on what they are learning, or need to learn, and how. One student raised the important point that if we assume future AI to be able to reliably generate code for whatever needed, then the coding skills needed will be "*about your ability to read and put together code as opposed to your ability to generate code*" (I1) or that students just need to "*fine-tune*" generated code. A similar view and concern is echoed by I3 who stated that "*I think it's a massive impact and one of the most pressing issues. I don't see how computer science at the university level can proceed in the same way. For example, I had a project this year that involved a significant amount of code development. If I had used AI, I could have completed it in a minute. This raises questions about the value of coding lessons and submissions*" (I3). However, most of the coding-specific comments were at a lower level focusing on testing, bug fixing, and problem-solving process.

#### 4.5.2 Testing

There were several examples from student on using GenAI to generate unit tests (e.g. "*…when it comes to unit testing. It's a very tedious and long process, which still needs to be done. So the use of GitHub Co-pilot is pretty good. I believe most of the group was using it to some extent towards the end*", I4; and "*It partially helped the team with writing test units by using our initial ideas to generate some tests using a new framework to the team.*", S19).

Another usage scenario was mentioned by I8 who explained how their group used GenAI to generate real-word testing data, which is another great potential for such tools. "*We used ChatGPT a lot for our test data… we wanted to use realistic data rather than a typical seeds generator, which would just put random text and random values into it. So, what we did was we gave ChatGPT all of our database schemas and we told it, 'Here's the relation between everything in our database. Can you generate some test data using real-world values?'*" (I8).

#### 4.5.3 Bug fixing

An expected use case with many examples is the use GenAI's at the bug-fixing level (e.g. "*when I added that code [GenAI generated code] in it worked. and it was just one line that I would never have thought to have put in my code.*", I5; "*I really only used ChatGPT to find bugs and errors in my code and not necessarily to write code.*", S16; "*A.I helped isolate and fix errors in code I had already written*" (S35); with many sharing the view that GenAI is a "*tool like Stackoverflow or Google that can assist a programming in writing safer and cleaner code.*" With some explaining that their reason for rating their experience with the use of AI positively because "*It's very easy to debug codes with ai*" (S1).

#### 4.5.4 Problem solving workflow

When it comes to the problem-solving workflow beyond simple bug fixing, and as with most other points, there were views on both sides of the spectrum. For some it completely changed how they solve problems (e.g. "*usually start with a Google search aiming for Stack Overflow. If that didn't seem to help, then I would go to the documentation. And then usually the final resort was actually asking a question on Stack Overflow myself... Whereas now, the main order of stuff is to ask Chat GPT if it can see any errors...*", I1; and "I would say the process of seeking help online have been reduced due to the use of AI tools within the project.", I6). While others still started their problem-solving process as they would have done pre-AI and then consulted AI if all failed and viewed GenAI as just another tool (e.g. "*I don't think it necessarily changed any of the initial steps I would have taken. Google is still often the easiest way to initially solve a problem. However, if I'm not entirely sure what's gone wrong, using AI tools can* *sometimes help...*", I8; and "*It didn't change my approach, only took the place of conventional websites such as stack overflow when the solution couldn't be found there.*", 23).

*4.5.5 Productivity and avoiding repetitive tasks.*

The data above provide multiple references to improving individual and team productivity weather in generating code and particularly repetitive code that they already know how to generate, debugging, problem solving, or brainstorming new ideas. This was a recurring theme not only at the team level but also at the individual level in both the interviews and surveys referencing its help in 'mundane' tasks (e.g. "*...It's a fantastic tool which allows us to remove a lot of the mundane or boilerplate code that you might have to write, no matter whether it's getters or setters and all that fun stuff....*", I8; " *So it saved a lot of like the manual hideous typing.*", I5; and "*Writing CSS classes can be tedious, and AI tools can quickly generate these, saving significant time.*",I7; and "*It was most helpful at completing code that was partially written and generally meant that features could be implemented quicker.*", S25).

Even when some students commented that AI can generate incorrect code, or not exactly what they wanted, they still found it to be useful, helpful and saves time "*.... Although sometimes it gave you the wrong answer or uncorrelated one. Focused on the technique which I'm not very familar, It also can help me a lot.*" (S26). But for others, it was just a waste of their time (e.g. "*it hindered my work as I tried to use it and it wasted my time*", S14; and "*I personally did not use chatgpt for coding because I every time I have it has been detrimental to my time spent programming. A quick google search is almost always better (for me)*", S33).

*4.5.6 Trust in AI code*

When discussing the topic of trust in AI generated code, there were varying views: Some were sceptics (e.g. "*I'll be honest. I don't trust it 100%. So I always, before I put it in, especially whenever we're doing group projects, I'll make my own branch, I'll put it in, I'll test it. I'll see if it causes any collisions anywhere else before I'll put it in.*", I2) while others were more trusting (e.g. "*...There were once or twice when I had my doubts, but I was pleasantly surprised as it still performed the actions exactly how I intended. So, I would say, it's pretty trustworthy.*", I4).

Several students stated that they started to trust it more during the project (e.g. "*I trusted it quite a bit from the start. It's been trained by Microsoft for years, and I started to trust it even more over time. For a university project, it's fully capable of helping out and writing very simple things.*", I4; "*I'd I'd say, it's yeah. my trust become more positive, I suppose, and just based on how that was.*", I6; and "*Over time, especially with the use of Chat GPT and the emergence of newer versions, my trust in AI has increased, particularly in coding projects.*", I7). But on the other end, we have one example of a student who started to trust it less stating, "*I think I initially trusted it much more than I do now.*" (I8).

## 5 INTERPRETATION AND CONTEXTUALIZATION OF RESULTS

### 5.1 Re-examining the data from different perspectives

Having presented our results through the lenses of teamwork, learning, coding specific uses, and productivity, it is useful to look at the same data from three other different perspectives: The roles that GenAI played during the project, the effects it had on the process and the learning, and finally a design perspective (i.e., what design dimensions can we deduce from the data and our analysis?).

Our results show the roles that GenAI played in this teamwork-based project can be classified into three distinct categories: As an educator, as a peer, and as an assistant.

- As an **educator**, the data from the learning theme shows that GenAI helped provide explanations, worked examples, alternative approaches to doing things, starting points, solve problems and even act as a mentor.
- As a **peer**, the data from the teamwork theme shows that GenAI played the role of a reviewer to code and to documentation and as a member contributing to brainstorming and idea generation.
- As an **assistant**, at the lowest level and from the coding-specific theme, we can see that GenAI played the role of an assistant helping write boilerplate code, writing unit testing, generating testing data, fixing bug, or turning small ideas into code.

Moreover, re-examining our data with a focus on the effects of the different roles that GenAI played, we can identify many positive effects but also potentially negative consequences, as illustrated in Table 3.

Table 3: The range of positive effects and potential negative consequences of GenAI as identified from the students' data.

| Positive effects of GenAI | Potential negative consequences of GenAI |
|---|---|
| <ul><li>Help focusing on:<ul><li>the bigger picture/idea.</li><li>the problem.</li><li>team skills.</li><li>testing.</li><li>design.</li></ul></li><li>Grasping the basics.</li><li>Learning new methods.</li><li>Building critical analysis skills.</li><li>Stretching students' abilities.</li><li>Bridging the skills gap.</li><li>Increasing students and teams' confidence (self/team-efficacy).</li><li>Reduced teams' anxiety and stress.</li><li>Not having to worry about syntax.</li><li>Improving productivity.</li></ul> | <ul><li>Slowing people down.</li><li>Hiding actual skills and knowledge.</li><li>Resulting in subpar code.</li><li>Resulting in working code that is not understood by students.</li><li>Generating code with bugs.</li><li>May disadvantage some students.</li><li>May negatively impact learning:<ul><li>not learning from mistakes.</li><li>not learning syntax.</li><li>learning the code, but not the underlying fundamentals.</li></ul></li></ul> |

It is interesting that there are some contradictory effects: improving productivity vs slowing people down, testing and bug fixing vs generating buggy code, stretching abilities vs overreliance thus not learning, bridging the skills gap vs disadvantaging some students. Our interpretation is that those who know programming fundamentals and know how to use AI will be able to gain most, if not all the mentioned benefits, but those who do not will see more of its negative consequences. We acknowledge that it is not easy to reconcile these contradicting effects, but we also argue that being aware of them is an important first step to taking best advantage of the potential while working on mitigating the possible negative consequences. We also hope that the design space we are proposing with its dimensions and constraints help in designing GenAI-based learning tools that maximizes the benefits and minimizes such possible negative effects.

**5.2 A design space for GenAI-based programming learning tools**

Whilst looking whether GitHub Copilot is an asset or liability in pair programming, Dakhel et al. [14] concluded that it can become an asset for experts, but a liability for novices. Evidence of this can be found in quotes such as "*with AI, we were coding at a higher level, using functions that we might not fully understand. We were editing the code on the fly and didn't have a specific way to verify it*" (I3). We argue that these tools can be an asset for learners as well if designed with pedagogical support in mind. As such, from a design perspective, we ask the following question. What are the pedagogical design considerations to take into account if we are to 'augment' [30] existing tools by providing an 'educational' wrapper layer based on what we have learned? While designing an interface or providing specific design recommendations is outside the scope of this paper, we aim to provide a first version of a 'design space' for such pedagogical wrappers building on ours and existing work.

One important dimension that was clear in the data is '**roles**' which currently include the three roles described earlier: educator, peer, and assistant. This dimension helps designers think of the type of support that they want to make available to learners when they interact with GenAI through such interfaces. Is the intention to limit and constrain GenAI support to the more mundane assistant roles, do we want to focus on the higher-level educational roles only, or do we want to leave it open without restrictions? Existing work that developed GenAI-powered tools such as WorkedGen on providing worked examples [23] and CodeHelp for providing explanation and feedback [33] fall under the educator support. Some aspects of the latter also fall under the assistant type, but we are not aware of any tools that utilize GenAI to specifically explore constraining support to be in the peer-support or assistant support categories.

An orthogonal dimension to that is '**support-ability patterns**' for each of role and their subsequent types of support (Figure 5). To understand how this dimension works with the type of support dimension, we can think of scaffolding and fading and how they can be applied to the educator role based on the student ability. As suggested by [26], in the beginning, the student may need a lot of support to start solving the problem and this support should decrease gradually (decreasing pattern: Figure 4a). However, in alignment with the guardrails idea used in [33], they also suggested ensuring the tool controlling over-utilization, which in their work involved ensuring not providing code in the answers as an example. We view these recommendations to fall under this dimension as they related to controlling the level of support as students' ability develops. Similarly, in the context of worked examples, support can start with full worked examples and a lot of explanations, then as students advance in their learning, the level of supports starts to fade away until it disappears (e.g. Abdul-Rahman and Boulay [1] refer to using partial examples or programming plans as examples of fading the scaffolding of worked-examples). On the other hand, in the capacity of an assistant role, and when it comes to providing mundane support such as syntax or unit testing, the educator may choose to not allow for AI support initially so novice students learn how to do these tasks, but as students ability develops and the focus shifts from syntax or testing to higher-level problem solving, the teacher may allow for increasing level of AI support (increasing pattern: Figure 4b). For roles such as peer support, we can imagine a certain constant level of AI support to be allowed regardless of students' abilities (constant pattern: Figure 4c).

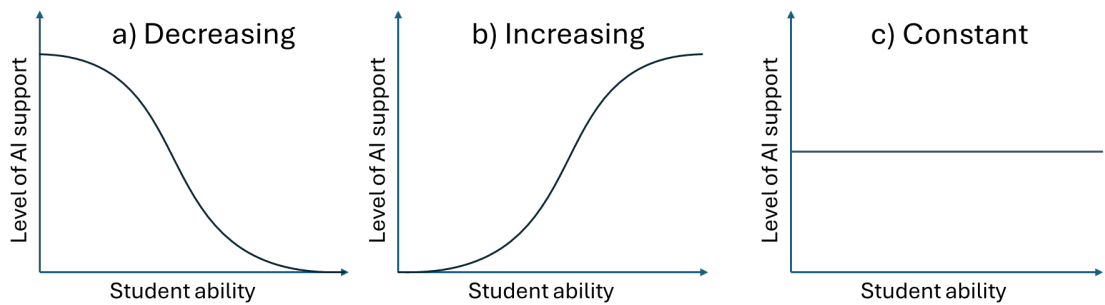

Figure 4: Example support-ability patterns. a) Level of AI support decreases as students abiity increases. b) Level of AI support increases as students' ability increases. c) Maintainint a constant level of AI support regardless of studets' ability levels.

The goal behind this dimension is to encourage designers to think about the level of support vs student ability level for each type of support intended. If GenAI is to be only an assistant, then what level and type of mundane tasks that it should provide and when? Should it immediately propose a fix to an identified bug, or should it guide the students to try and find the bug and fix it, should it provide support as a starting point, or later in the process? While the support-ability patterns for specific types of support within each role can be simple increasing, decreasing, or constant, the overall combined level of AI support required may end up being shaped like a bell-curve, U-shaped curve or more a more complex shape. Ideally each type of support should be adaptive to each student such that the provided AI support shifts or extend each student's zone of proximal development [50] allowing them to tackle more difficult learning tasks than they would without AI support. Figure 5 illustrates this by applying the concept of ZPD [50] to that of Flow [35] to demonstrate how choosing the right support-ability patterns for the different types of support and students ability level has the potential to shift this zone towards more difficult tasks.

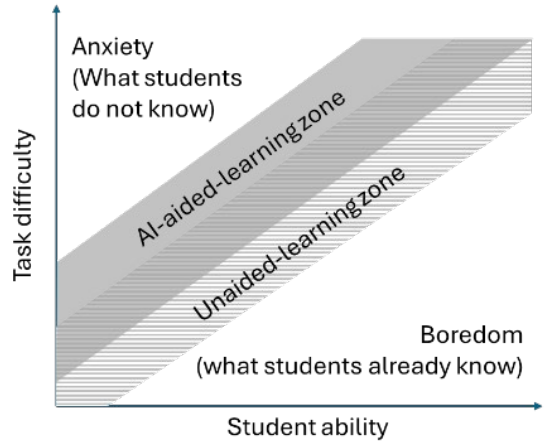

Figure 5: AI support has the potential to shift students' zone of proximal development to allow them to tackle more challenging tasks.

The third dimension is the '**transparency**' dimension. Designers need to consider how the tool can ensure a high level of transparency as to the way, and the extent, to which AI was used in the learning process. Designers can consider features like interaction history which communicate the learning process and not just the output, digital watermarks embedded in the students' submission, or any other features that ensure transparency as to how GenAI was used in the learning process. Supporting teachers to observe, summarize, and review how students engage in use of GenAI tool was provided in CodeHelp [33] but these features were for educators only. Our results show that students also expressed concerns that the use of GenAI tools can 'hide' the real skills of the students from both the educators and their team members. As such, we view this as in important dimension to consider not only for educators and the purpose of assessment, but also for team members and throughout all the stages of work. Even providing it for the students' themselves may show them that they are over-relying on such tools and encourage some degree of self-regulation as to how, and to what extent, such tools are used.

With careful consideration to these three dimensions, the different potential negative consequences of GenAI mentioned earlier can be addressed whether they are about hiding skills (e.g. increased transparency), not learning from mistakes (e.g. ensuring support for an educational role), or not learning syntax or fundamentals (e.g. using a support pattern that restrains the level and type of support provided for novice users). For each student/team and at each level in their learning journey and depending on the desired goals, there will be a region in this 3D design space that can maximise the learning benefits and improve the learning experience.

## 6 DISCUSSION

In this work, we present and analyze findings from interview and survey responses from students who took a software engineering course with the goal of gaining students' informed insights on the potential role and implications of GenAI in software engineering education. The module used a 7-week group project-based assessment, and students were explicitly given permission to use GenAI tools to help with the coding part of their project.

The rich qualitative data we were able to gain from the students helps enrich our understanding of the role of GenAI in education as viewed from the students' perspectives based on their actual use in their team project (first contribution). As this was a group-based project, the students' comments and answers included new insights as to the roles that GenAI played within a teamwork-based context (second contribution). Our current understanding of GenAI's collaborative role in computing education is mainly focused on human-AI collaboration in settings such as pair programming with AI (e.g. [6, 34, 54]) rather than AI being another member of a larger team. Our findings show that GenAI can play a role not just as an assistant or an educator, but that of a team member. In this role, GenAI was a peer that influenced team decisions and outcomes in activities such as ideation, brainstorming, and for reviewing code and documentation. It also showed how some team members viewed GenAI as bridging the skills gap within their teams, thus increasing the confidence of team members and consequently of the team. GenAI was also perceived to improve the productivity of the team, helping shift its focus to the bigger picture, team skills and testing. At the same time, there were cases where GenAI's usage within the team lead to some challenges like students contributing unchecked buggy, or subpar code that was generated by AI without them understanding it. Some students also raised concerns about GenAI hiding the real ability level of other team members which raises concerns around accountability which can negatively impact the team dynamics.

The third contribution of this work is the use of learning theory and pedagogy lens to examine the students' input. This perspective helped us identify sub-themes such as the role of GenAI in providing general scaffolding and fading [7] and providing support within some students' Zone of Proximal Development [50] even in its current ChatGPT or Copilot format. Other sub-themes that fall under different approaches to scaffolding include using GenAI generated code as worked examples (also explored in [23, 32]), scaffolding in getting started with a problem, and in learning different ways of doing things. The latter point is important not only in developing students general coding skills but also their critical thinking skills when it comes to judging the pros and cons of different approaches to achieve the same goal. Just as GenAI helped improve group-efficacy, it also improved individual students' self-efficacy and their confidence in that they can solve problems they would not have tried to solve before. A main conclusion from all the input whether it is on the positive role or negative implication is that it all depends on when help from GenAI is sought and to what extent. Students stressed the importance of understanding the fundamentals first to be able to benefit from GenAI, avoid over-reliance and ensure it does not affect their learning. Such emphasis on mastering the fundamentals first keeps emerging whether in talking to professional developers, educators or students (e.g. [11, 33, 41]).

The fourth and final contribution of this work is the proposal of a preliminary design space for designing GenAI based programming learning tools/environments. Through a multiple perspective analysis of our data including a learning and pedagogy perspective, roles perspective, and benefits and consequences perspectives, we were able to identify some dimensions that are important to consider by any educator or designer looking to develop an educational layer around such tools (e.g. [26, 33]). Our design space includes looking at the **roles** that GenAI can/should support such as educator, peer, or assistant with the functionalities that can be supported in each. It also includes looking at the desired patterns of **support-ability** level for each role and for each functionality within these roles. Examples of such support-ability patterns include delayed support for novices for certain aspects until achieving a certain level of knowledge. This contrasts with a faded support pattern that starts at a maximum for novices but fades with increased level of knowledge. However, in some cases, a simple constant level of support can work best. These patterns can be reapplied for each new topic and are not expected to apply for the whole duration of a module. The final dimension of our design space is that of **transparency**. This aims to ensure that any wrapper around existing GenAI tools provides enough transparency not just to educators but also to peers and the individuals themselves as to the level and nature of use of GenAI to help in providing formative feedback, assessment, accountability, and self-reflection. CodeHelp [33] provides an example of how this can be done for the educators, but more exploration is needed as to how this can be done effectively at the team and individual levels.

We hope that this early exploration into a design space for GenAI-powered computing education tools provides a more structured approach to the design and development of such tools and that this work can be extended or adapted to other contexts beyond computing education. While some of the functionalities in the roles dimension are discipline dependent, these can be replaced by other discipline-specific functionalities. The other dimensions are more abstract which means after identifying discipline-specific roles and functionalities, the whole design space can be easily adapted and adopted to other disciplines.

More generally, many of the themes, subthemes, positive effects and potential negative consequences identified from the students experience in this computing module can provide general valuable insights to educators including the impact on teamwork, learning, group and self-efficacy, and productivity. As such, we hope that this work and its contributions extend beyond the field of computing education and help enrich the findings and discussions of this quickly evolving area of research.

## 7  LIMITATIONS AND FUTURE WORK

Despite the incentives provided, only 39 students responded to our survey and just eight agreed to participate in the follow-up semi-structured interviews. While this sample resulted in rich qualitative and varied responses, a larger sample size may offer further insights. Additionally, the proposed design space has not been validated. Our aim is for this to be viewed a starting point to encourage discussion and research to develop this design space further including expanding the possible roles that GenAI can play in different contexts, detailed investigation into the different possible support-ability patterns, as well as refining and adding to the proposed dimensions. This work calls for further exploration of GenAI's role into teamwork context, as ours is just one case study in this space. We advocate for going beyond merely reporting students and educators' views, suggesting the use of appropriate learning theories and pedagogical lenses in analysis and reporting. Finally, we encourage more design-focused exploration of this rapidly evolving technology to enhance its educational potential and address its associated concerns.

## 8 CONCLUSION

This study provides valuable insights into the integration of generative AI tools, such as ChatGPT and GitHub Copilot, in undergraduate software engineering education and particularly in teamwork-based projects. By analyzing students' experiences and perspectives, we uncovered both the benefits and challenges of using GenAI in team-based coding projects. Our findings emphasize GenAI's potential positive impact on a variety of aspects such as learning and learning experience, productivity, individual and team efficacy, and team dynamics. At the same time, the findings also highlight the need for mindful implementation to mitigate different or complementary issues including over-reliance, accountability, and negative impact on learning. The proposed design space offers a starting point for developing a structured framework for guiding the development GenAI-based educational tools to maximize the identified benefits and mitigate the possible negative consequences by considering varied roles, support-ability patterns, and transparency. These insights contribute to a deeper understanding of GenAI's role in computing education in particular, but we also discussed how these can be generalized to other disciplines. The novel contributions of this work lie in its emphasis on GenAI's role within teamwork, the use of learning theory and pedagogical lenses to discuss the learning implications, and the proposed design space to inform the design and application of GenAI within educational contexts.

**ACKNOWLEDGMENTS**

This research was funded by the Center for Digital Citizens (EP/T022582/1).